\documentclass[fleqn]{llncs}
\usepackage{version}
\usepackage{tpis2}

\title{A Protocol for Instruction Stream Processing}
\author{J.A. Bergstra \and C.A. Middelburg}
\institute{Informatics Institute, Faculty of Science,
           University of Amsterdam, \\
           Science Park~107, 1098~XG Amsterdam, the Netherlands \\
           \email{J.A.Bergstra@uva.nl,C.A.Middelburg@uva.nl}}

\begin{document}

\maketitle

\begin{abstract}
The behaviour produced by an instruction sequence under execution is a
behaviour to be controlled by some execution environment: each step
performed actuates the processing of an instruction by the execution
environment and a reply returned at completion of the processing
determines how the behaviour proceeds.
In this paper, we are concerned with the case where the processing takes
place remotely.
We describe a protocol to deal with the case where the behaviour
produced by an instruction sequence under execution leads to the
generation of a stream of instructions to be processed and a remote
execution unit handles the processing of that stream of instructions.
\begin{classcode}
D.2.1, D.2.4, F.1.1, F.3.1.
\end{classcode}
\end{abstract}

\section{Introduction}
\label{sect-intro}

The behaviour produced by an instruction sequence under execution is a
behaviour to be controlled by some execution environment.
It proceeds by performing steps in a sequential fashion.
Each step performed actuates the processing of an instruction by the
execution environment.
A reply returned by the execution environment at completion of the
processing of the instruction determines how the behaviour proceeds.
Often, the processing of instructions takes place remotely.
This means that, on execution of an instruction sequence, a stream of
instructions to be processed arises at one place and the processing of
that stream of instructions is handled at another place.
The objective of the current paper is to bring this phenomenon better
into the picture.
To achieve this objective, we describe a basic protocol for instruction
stream processing that deals with this phenomenon.

The phenomenon sketched above is found if it is impracticable to load the
instruction sequence to be executed as a whole.
For instance, the storage capacity of the execution unit is too small or
the execution unit is too far away.
The phenomenon requires special attention because the transmission time
of the messages involved in remote processing makes it hard to keep the
execution unit busy without intermission.
The basic protocol for instruction stream processing is directed towards
keeping the execution unit busy.
There is no reason to use the word ``remote'' in a narrow sense.
It is convenient to consider processing remote if it involves message
passing with transmission times that are not negligible.
In that case, the basic protocol provides a starting-point for studies
of basic techniques aimed at increasing processor performance, such as
pre-fetching and branch-prediction, at a more abstract level than usual.
Therefore, we consider the protocol relevant to the area of computer
architectures.

The work presented in this paper is a spin-off of the line of research
with which a start was made in~\cite{BL02a}.
The working hypothesis of that line of research is that instruction
sequence is a central notion of computer science.
In that line of research, issues concerning the following subjects from
the theory of computation have been investigated from the viewpoint that
a program is an instruction sequence: semantics of programming
languages, expressiveness of programming languages, computability,
computational complexity, and performance related matters of instruction
sequences (see e.g.~\cite{BM07e,BM07g,BP04a,BM08g,BM07f}).
The description of the basic protocol for instruction stream processing
starts from the behaviours produced by instruction sequences under
execution.
By that we abstract from the instruction sequences which produce those
behaviours.
At the level of abstraction concerned, it is easy to describe how the
instruction streams are generated.
How instruction streams can be generated efficiently from instruction
sequences is another matter.

Threads as considered in \BTA\ (Basic Thread Algebra) are used in this
paper to model the behaviours produced by instruction sequences under
execution.
\BTA, introduced under the name \BPPA\ (Basic Polarized Process Algebra)
in~\cite{BL02a}, is a form of process algebra tailored to the
description and analysis of the behaviours produced by instruction
sequences under execution.
General process algebras, such as \ACP~\cite{BK84b,BW90},
CCS~\cite{HM85,Mil89} and CSP~\cite{BHR84,Hoa85}, are too general for
the description and analysis of the behaviours produced by instruction
sequences under execution.
That is, it is quite awkward to describe and analyse behaviours of this
kind using such a general process algebra.
However, the behaviours considered in \BTA\ can be viewed as processes
that are definable over \ACP, see e.g.~\cite{BM05c}.
This allows for the basic protocol for instruction stream processing to
be described using \ACP\ or rather \ACPt, an extension of \ACP\ which
supports abstraction from internal actions.

This paper is organized as follows.
First, we give brief summaries of \BTA\ (Section~\ref{sect-BTA}) and
\ACPt\ (Section~\ref{sect-ACP}).
Next, we show how the behaviours considered in \BTA\ can be viewed as
processes that are definable over \ACPt\
(Section~\ref{sect-process-extr}).
Then, we describe the basic protocol for instruction stream processing
(Section~\ref{sect-protocol}) and discuss some conceivable adaptations
of the protocol (Section~\ref{sect-adaptations}).
Finally, we make some concluding remarks (Section~\ref{sect-concl}).

\section{Thread Algebra}
\label{sect-BTA}

In this section, we review \BTA\ (Basic Thread Algebra).
\BTA\ is concerned with behaviours as exhibited by instruction sequences
under execution.
These behaviours are called threads.

In \BTA, it is assumed that a fixed but arbitrary set $\BAct$ of
\emph{basic actions} has been given.
A thread performs basic actions in a sequential fashion.
Upon each basic action performed, a reply from the execution environment
of the thread determines how it proceeds.
The possible replies are the Boolean values $\True$ and~$\False$.

To build terms, \BTA\  has the following constants and operators:
\begin{iteml}
\item
the \emph{deadlock} constant $\DeadEnd$;
\item
the \emph{termination} constant $\Stop$;
\item
for each $a \in \BAct$, the binary \emph{postconditional composition}
operator $\pccop{a}$.
\end{iteml}
We assume that there are infinitely many variables, including $x,y,z$.
Terms are built as usual.
We use infix notation for the postconditional composition operator.
We introduce \emph{basic action prefixing} as an abbreviation:
$a \bapf p$, where $a \in \BAct$ and $p$ is a \BTA\ term, abbreviates
$\pcc{p}{a}{p}$.

The thread denoted by a closed term of the form $\pcc{p}{a}{q}$ will
first perform $a$, and then proceed as the thread denoted by $p$
if the reply from the execution environment is $\True$ and proceed as
the thread denoted by $q$ if the reply from the execution environment is
$\False$.
The threads denoted by $\DeadEnd$ and $\Stop$ will become inactive and
terminate, respectively.
This implies that each closed \BTA\ term denotes a thread that will
become inactive or terminate after it has performed finitely many
basic actions.
Infinite threads can be described by guarded recursion.

A \emph{guarded recursive specification} over \BTA\ is a set of
recursion equations $E = \set{X = t_X \where X \in V}$, where $V$ is a
set of variables and each $t_X$ is a \BTA\ term of the form $\DeadEnd$,
$\Stop$ or $\pcc{t}{a}{t'}$ with $t$ and $t'$ that contain only
variables from $V$.
We write $\vars(E)$ for the set of all variables that occur in $E$.
We are only interested in models of \BTA\ in which guarded recursive
specifications have unique solutions, such as the projective limit model
of \BTA\ presented in~\cite{BB03a}.

For each guarded recursive specification $E$ and each $X \in \vars(E)$,
we introduce a constant $\rec{X}{E}$ standing for the unique solution of
$E$ for $X$.
The axioms for these constants are given in Table~\ref{axioms-rec}.%
\begin{table}[!t]
\caption{Axioms for guarded recursion}
\label{axioms-rec}
\begin{eqntbl}
\begin{saxcol}
\rec{X}{E} = \rec{t_X}{E} & \mif X \!=\! t_X \in E       & \axiom{RDP}
\\
E \Implies X = \rec{X}{E} & \mif X \in \vars(E)          & \axiom{RSP}
\end{saxcol}
\end{eqntbl}
\end{table}
In this table, we write $\rec{t_X}{E}$ for $t_X$ with, for all
$Y \in \vars(E)$, all occurrences of $Y$ in $t_X$ replaced by
$\rec{Y}{E}$.
RDP and RSP are actually axiom schemas in which $X$ stands for an
arbitrary variable, $t_X$ stands for an arbitrary \BTA\ term, and $E$
stands for an arbitrary guarded recursive specification over \BTA.
Side conditions are added to restrict what $X$, $t_X$ and $E$ stand for.

Let $\fM$ be a model of \BTA\ extended with guarded recursion.
Then we use the term thread for the elements from the domain of $\fM$,
and we denote the interpretations of constants and operators in $\fM$ by
the constants and operators themselves.
Moreoever, let $p$ be a thread.
Then the set of \emph{states} or \emph{residual threads} of $p$,
written $\Res(p)$, is inductively defined as follows:
\begin{itemize}
\item
$p \in \Res(p)$;
\item
if $\pcc{q}{a}{r} \in \Res(p)$, then $q \in \Res(p)$ and
$r \in \Res(p)$.
\end{itemize}
We say that $p$ is a \emph{regular} thread if $\Res(p)$ is finite.
Being a regular thread coincides with being the solution of a finite
guarded recursive specification.

In the sequel, we will make use of a version of \BTA\ in which the
following additional assumptions relating to $\BAct$ are made:
(i)~a fixed but arbitrary set $\Foci$ of \emph{foci} has been given;
(ii)~a fixed but arbitrary set $\Meth$ of \emph{methods} has been given;
(iii)~$\BAct = \set{f.m \where f \in \Foci, m \in \Meth}$.
These assumptions are based on the view that the execution environment
provides a number of services.
Performing a basic action $f.m$ is taken as making a request to the
service named $f$ to process command $m$.
As usual, we will write $\Bool$ for the set $\set{\True,\False}$.

\section{Process Algebra}
\label{sect-ACP}

In this section, we review \ACPt\ (Algebra of Communicating Processes
with abstraction).
This is the process algebra that will be used in
Section~\ref{sect-process-extr} to make precise what processes are
produced by the threads denoted by closed terms of \BTA\ with guarded
recursion.
For a comprehensive overview of \ACPt, the reader is referred
to~\cite{BW90,Fok00}.

In \ACPt, it is assumed that a fixed but arbitrary set $\Act$ of
\emph{atomic actions}, with $\tau,\dead \notin \Act$, and a fixed but
arbitrary commutative and associative function
$\funct{\commm}{\Act \x \Act}{\Act \union \set{\dead}}$
have been given.
The function $\commm$ is regarded to give the result of synchronously
performing any two atomic actions for which this is possible, and to
give $\dead$ otherwise.
In \ACPt, $\tau$ is a special atomic action, called the silent step.
The act of performing the silent step is considered unobservable.
Because it would otherwise be observable, the silent step is considered
an atomic action that cannot be performed synchronously with other
atomic actions.

\ACPt\ has the following constants and operators:
\begin{itemize}
\item
for each $e \in \Act$, the \emph{atomic action} constant $e$\,;
\item
the \emph{silent step} constant $\tau$\,;
\item
the \emph{deadlock} constant $\dead$\,;
\item
the binary \emph{alternative composition} operator $\altc$\,;
\item
the binary \emph{sequential composition} operator $\seqc$\,;
\item
the binary \emph{parallel composition} operator $\parc$\,;
\item
the binary \emph{left merge} operator $\leftm$\,;
\item
the binary \emph{communication merge} operator $\commm$\,;
\item
for each $H \subseteq \Act$, the unary \emph{encapsulation} operator
$\encap{H}$\,;
\item
for each $I \subseteq \Act$, the unary \emph{abstraction} operator
$\abstr{I}$\,.
\end{itemize}
We assume that there are infinitely many variables, including $x,y,z$.
Terms are built as usual.
We use infix notation for the binary operators.

Let $p$ and $q$ be closed \ACPt\ terms, $e \in \Act$, and
$H,I \subseteq \Act$.
Intuitively, the constants and operators to build \ACPt\ terms can be
explained as follows:
\begin{itemize}
\item
$e$ first performs atomic action $e$ and next terminates successfully;
\item
$\tau$ performs an unobservable atomic action and next terminates
successfully;
\item
$\dead$ can neither perform an atomic action nor terminate successfully;
\item
$p \altc q$ behaves either as $p$ or as $q$, but not both;
\item
$p \seqc q$ first behaves as $p$ and on successful termination of $p$
it next behaves as~$q$;
\item
$p \parc q$ behaves as the process that proceeds with $p$ and $q$ in
parallel;
\item
$p \leftm q$ behaves the same as $p \parc q$, except that it starts
with performing an atomic action of $p$;
\item
$p \commm q$ behaves the same as $p \parc q$, except that it starts with
performing an\linebreak[2] atomic action of $p$ and an atomic action of
$q$ synchronously;
\item
$\encap{H}(p)$ behaves the same as $p$, except that atomic actions from
$H$ are blocked;
\item
$\abstr{I}(p)$ behaves the same as $p$, except that atomic actions from
$I$ are turned into unobservable atomic actions.
\end{itemize}
%

The axioms of \ACPt\ are given in Table~\ref{axioms-ACPt}.
\begin{table}[!t]
\caption{Axioms of \ACPt}
\label{axioms-ACPt}
\begin{eqntbl}
\begin{axcol}
x \altc y = y \altc x                                  & \axiom{A1}  \\
(x \altc y) \altc z = x \altc (y \altc z)              & \axiom{A2}  \\
x \altc x = x                                          & \axiom{A3}  \\
(x \altc y) \seqc z = x \seqc z \altc y \seqc z        & \axiom{A4}  \\
(x \seqc y) \seqc z = x \seqc (y \seqc z)              & \axiom{A5}  \\
x \altc \dead = x                                      & \axiom{A6}  \\
\dead \seqc x = \dead                                  & \axiom{A7}  \\
{}                                                                   \\
x \parc y =
          x \leftm y \altc y \leftm x \altc x \commm y & \axiom{CM1} \\
a \leftm x = a \seqc x                                 & \axiom{CM2} \\
a \seqc x \leftm y = a \seqc (x \parc y)               & \axiom{CM3} \\
(x \altc y) \leftm z = x \leftm z \altc y \leftm z     & \axiom{CM4} \\
a \seqc x \commm b = (a \commm b) \seqc x              & \axiom{CM5} \\
a \commm b \seqc x = (a \commm b) \seqc x              & \axiom{CM6} \\
a \seqc x \commm b \seqc y =
                        (a \commm b) \seqc (x \parc y) & \axiom{CM7} \\
(x \altc y) \commm z = x \commm z \altc y \commm z     & \axiom{CM8} \\
x \commm (y \altc z) = x \commm y \altc x \commm z     & \axiom{CM9}
\end{axcol}
\qquad
\begin{axcol}
x \seqc \tau = x                                       & \axiom{B1}  \\
x \seqc (\tau \seqc (y \altc z) \altc y) = x \seqc (y \altc z)
                                                       & \axiom{B2}  \\
{}                                                                   \\
\encap{H}(a) = a                \hfill \mif a \notin H & \axiom{D1}  \\
\encap{H}(a) = \dead            \hfill \mif a \in H    & \axiom{D2}  \\
\encap{H}(x \altc y) = \encap{H}(x) \altc \encap{H}(y) & \axiom{D3}  \\
\encap{H}(x \seqc y) = \encap{H}(x) \seqc \encap{H}(y) & \axiom{D4}  \\
{}                                                                   \\
\abstr{I}(a) = a                \hfill \mif a \notin I & \axiom{TI1} \\
\abstr{I}(a) = \tau             \hfill \mif a \in I    & \axiom{TI2} \\
\abstr{I}(x \altc y) = \abstr{I}(x) \altc \abstr{I}(y) & \axiom{TI3} \\
\abstr{I}(x \seqc y) = \abstr{I}(x) \seqc \abstr{I}(y) & \axiom{TI4} \\
{}                                                                   \\
a \commm b = b \commm a                                & \axiom{C1}  \\
(a \commm b) \commm c = a \commm (b \commm c)          & \axiom{C2}  \\
\dead \commm a = \dead                                 & \axiom{C3}  \\
\tau \commm a = \dead                                  & \axiom{C4}
\end{axcol}
\end{eqntbl}
\end{table}
CM2--CM3, CM5--CM7, C1--C4, D1--D4 and TI1--TI4 are actually axiom
schemas in which $a$, $b$ and $c$ stand for arbitrary constants of
\ACPt, and $H$ and $I$ stand for arbitrary subsets of $\Act$.
\ACPt\ is extended with guarded recursion like \BTA.

A \emph{recursive specification} over \ACPt\ is a set of recursion
equations $E = \set{X = t_X \where X \in V}$, where $V$ is a set of
variables and each $t_X$ is an \ACPt\ term containing only variables
from $V$.
We write $\vars(E)$ for the set of all variables that occur in $E$.
Let $t$ be an \ACPt\ term without occurrences of abstraction operators
containing a variable $X$.
Then an occurrence of $X$ in $t$ is \emph{guarded} if $t$ has a subterm
of the form $e \seqc t'$ where $e \in \Act$ and $t'$ is a term
containing this occurrence of $X$.
Let $E$ be a recursive specification over \ACPt.
Then $E$ is a \emph{guarded recursive specification} if, in each
equation $X = t_X \in E$:
(i)~abstraction operators do not occur in $t_X$ and
(ii)~all occurrences of variables in $t_X$ are guarded or $t_X$ can be
rewritten to such a term using the axioms of \ACPt\ in either direction
and/or the equations in $E$ except the equation $X = t_X$ from left to
right.
We are only interested models of \ACPt\ in which guarded recursive
specifications have unique solutions, such as the models of \ACPt\
presented in~\cite{BW90}.

For each guarded recursive specification $E$ and each $X \in \vars(E)$,
we introduce a constant $\rec{X}{E}$ standing for the unique solution of
$E$ for $X$.
The axioms for these constants are given in
Table~\ref{axioms-RDP-RSP}.%
\begin{table}[!t]
\caption{Axioms for guarded recursion}
\label{axioms-RDP-RSP}
\begin{eqntbl}
\begin{saxcol}
\rec{X}{E} = \rec{t_X}{E} & \mif X \!=\! t_X \in E         & \axiom{RDP}
\\
E \Implies X = \rec{X}{E} & \mif X \in \vars(E)            & \axiom{RSP}
\end{saxcol}
\end{eqntbl}
\end{table}
In this table, we write $\rec{t_X}{E}$ for $t_X$ with, for all
$Y \in \vars(E)$, all occurrences of $Y$ in $t_X$ replaced by
$\rec{Y}{E}$.
RDP and RSP are actually axiom schemas in which $X$ stands for an
arbitrary variable, $t_X$ stands for an arbitrary \ACPt\ term, and $E$
stands for an arbitrary guarded recursive specification over \ACPt.
Side conditions are added to restrict what $X$, $t_X$ and $E$ stand for.

We will write $\vAltc{i \in S} p_i$, where $S = \set{i_1,\ldots,i_n}$
and $p_{i_1},\ldots,p_{i_n}$ are \ACPt\ terms,
for $p_{i_1} \altc \ldots \altc p_{i_n}$.
The convention is that $\vAltc{i \in S} p_i$ stands for $\dead$ if
$S = \emptyset$.
We will often write $X$ for $\rec{X}{E}$ if $E$ is clear from the
context.
It should be borne in mind that, in such cases, we use $X$ as a
constant.

\section{Process Extraction}
\label{sect-process-extr}

In this section, we use \ACPt\ with guarded recursion to make
mathematically precise what processes are produced by the threads
denoted by closed terms of \BTA\ with guarded recursion.

For that purpose, $\Act$ and $\commm$ are taken such that the following
conditions are satisfied:
\begin{ldispl}
\begin{aeqns}
\Act & \supseteq &
\set{\snd_f(d) \where f \in \Foci, d \in \Meth \union \Bool} \union
\set{\rcv_f(d) \where f \in \Foci, d \in \Meth \union \Bool}
\union
\set{\stp,\iact}
\end{aeqns}
\end{ldispl}%
and for all $f \in \Foci$, $d \in \Meth \union \Bool$, and
$e \in \Act$:
\begin{ldispl}
\begin{aeqns}
\snd_f(d) \commm \rcv_f(d) = \iact\;,
\\
\snd_f(d) \commm e = \dead & & \mif e \neq \rcv_f(d)\;,
\\
e \commm \rcv_f(d) = \dead & & \mif e \neq \snd_f(d)\;,
\end{aeqns}
\qquad\;
\begin{aeqns}
{} \\
\stp \commm e = \dead\;,
\\
\iact \commm e = \dead\;.
\end{aeqns}
\end{ldispl}

The \emph{process extraction} operation $\pextr{\ph}$ determines, for
each closed term $p$ of \BTA\ with guarded recursion, a closed term of
\ACPt\ with guarded recursion that denotes the process produced by the
thread denoted by $p$.
The process extraction operation $\pextr{\ph}$ is defined by
$\pextr{p} = \abstr{\set{\stp}}(\cpextr{p})$, where $\cpextr{\ph}$ is
defined by the equations given in Table~\ref{eqns-process-extr}
(for $f \in \Foci$ and $m \in \Meth$).%
\begin{table}[!t]
\caption{Defining equations for process extraction operation}
\label{eqns-process-extr}
\begin{eqntbl}
\begin{eqncol}
\cpextr{X} = X
\\
\cpextr{\Stop} = \stp
\\
\cpextr{\DeadEnd} = \iact \seqc \dead
\\
\cpextr{\pcc{t_1}{f.m}{t_2}} =
\snd_f(m) \seqc
(\rcv_f(\True) \seqc \cpextr{t_1} \altc
 \rcv_f(\False) \seqc \cpextr{t_2})
\\
\cpextr{\rec{X}{E}} =
\rec{X}{\set{Y = \cpextr{t_Y} \where Y = t_Y \,\in\, E}}
\end{eqncol}
\end{eqntbl}
\end{table}

Two atomic actions are involved in performing a basic action of the form
$f.m$: one for sending a request to process command $m$ to the service
named $f$ and another for receiving a reply from that service upon
completion of the processing.
For each closed term $p$ of \BTA\ with guarded recursion, $\cpextr{p}$
denotes a process that in the event of termination performs a special
termination action just before termination.
Abstracting from this termination action yields the process denoted by
$\pextr{p}$.
Some atomic actions introduced above are not used in the definition of
the process extraction operation for \BTA.
Those atomic actions are commonly used in the definition of the process
extraction operation for extensions of \BTA\ in which operators for
thread-service interaction occur, see e.g.~\cite{BM05c}.

Let $p$ be a closed term of \BTA\ with guarded recursion.
Then we say that $\pextr{p}$ is the \emph{process produced by} $p$.

\sloppy
The process extraction operation preserves the axioms of \BTA\ with
guarded recursion.
Roughly speaking, this means that the translations of these axioms are
derivable from the axioms of \ACPt\ with guarded recursion.
Before we make this fully precise, we have a closer look at the axioms
of \BTA\ with guarded recursion.

A proper axiom is an equation or a conditional equation.
In Table~\ref{axioms-rec}, we do not find proper axioms.
Instead of proper axioms, we find axiom schemas without side conditions
and axiom schemas with side conditions.
The axioms of \BTA\ with guarded recursion are obtained by replacing
each axiom schema by all its instances.

We define a function $\transl{\ph}$ from the set of all equations and
conditional equations of \BTA\ with guarded recursion to the set of all
equations of \ACPt\ with guarded recursion as follows:
\begin{ldispl}
\transl{t_1 = t_2} \;\;=\;\; \pextr{t_1} = \pextr{t_2}\;,
\\
\transl{E \Implies t_1 = t_2} \;\;=\;\;
\set{\pextr{t'_1} = \pextr{t'_2} \where t'_1 = t'_2 \,\in\, E} \Implies
\pextr{t_1} = \pextr{t_2}\;.
\end{ldispl}
\begin{proposition}
\label{prop-preservation-axioms}
Let $\phi$ be an axiom of \BTA\ with guarded recursion.
Then $\transl{\phi}$ is derivable from the axioms of \ACPt\ with guarded
recursion.
\end{proposition}
\begin{proof}
The proof is trivial.
\qed
\end{proof}
Proposition~\ref{prop-preservation-axioms} would go through if no
abstraction of the above-mentioned special termination action was made.
Notice further that \ACPt\ without the silent step constant and the
abstraction operator, better known as \ACP, would suffice if no
abstraction of the special termination action was made.

\section{A Protocol for Instruction Stream Processing}
\label{sect-protocol}

In this section, we consider a protocol for instruction stream
processing.
Before the protocol is described, an extension of \ACP\ is introduced
to simplify the description of the protocol.

The following extension of \ACP\ from~\cite{BB92c} will be used below:
the non-branching conditional operator $\gc$ over
$\Bool = \set{\True,\False}$.
The expression $b \gc p$, is to be read as
\texttt{if} $b$ \texttt{then} $p$ \texttt{else} $\dead$.
The axioms for the non-branching conditional operator are
\begin{ldispl}
\True \gc x = x \quad \mathrm{and} \quad \False \gc x = \dead\;.
\end{ldispl}

The protocol concerns a system whose main components are an
\emph{instruction stream generator} and an \emph{instruction stream
execution unit}.
The instruction stream generator generates different instruction
streams for different threads.
This is accomplished by starting it in different states.
The general idea of the protocol is that:
\begin{itemize}
\item
the instruction stream generator generating an instruction stream for a
thread of the form $\pcc{t}{a}{t'}$ sends $a$ to the instruction stream
execution unit;
\item
on receipt of $a$, the instruction stream execution unit gets the
execution of $a$ done and sends the reply produced to the instruction
stream generator;
\item
on receipt of the reply, the instruction stream generator proceeds with
generating an instruction stream for $t$ if the reply is $\True$ and for
$t'$ otherwise.
\end{itemize}
In the case where the thread is $\Stop$ or $\DeadEnd$, the instruction
stream generator sends a special instruction ($\stopd$ or $\deadd$) and
the instruction stream execution unit does not send back a reply.
The specifics of the protocol considered here are that:
\begin{itemize}
\item
the instruction stream generator may run ahead of the instruction stream
execution unit by not waiting for the receipt of the replies resulting
from the execution of instructions that it has sent earlier;
\item
to ensure that the instruction stream execution unit can handle the
run-ahead, each instruction sent by the instruction stream generator is
accompanied with the sequence of replies after which the instruction
must be executed;
\item
to correct for replies that have not yet reached the instruction stream
generator, each instruction sent is also accompanied with the number of
replies received since the last sending of an instruction.
\end{itemize}

We write $\BActi$ for the set $\BAct \union \set{\stopd,\deadd}$.
Elements from $\BActi$ will loosely be called instructions.

Henceforth, it is assumed that a model of \BTA\ extended with guarded
recursion has been given.
The restriction of the domain of that model to the regular threads will
be denoted by $\Reg$.

The functions $\nm{act}$, $\nm{thrt}$, and $\nm{thrf}$ defined below
give, for each thread $t$ different from $\Stop$ and $\DeadEnd$, the
action that $t$ will perform first, the thread with which it will
proceed if the reply from the execution environment is $\True$, and the
thread with which it will proceed if the reply from the execution
environment is $\False$, respectively.
The functions $\funct{\nm{act}}{\Reg}{\BActi}$,
$\funct{\nm{thrt}}{\Reg}{\Reg}$, and $\funct{\nm{thrf}}{\Reg}{\Reg}$ are
defined as follows:
\begin{ldispl}
\begin{geqns}
\act{\Stop} = \stopd\;,
\\
\act{\DeadEnd} = \deadd\;,
\\
\act{\pcc{t}{a}{t'}} = a\;,
\end{geqns}
\quad
\begin{geqns}
\rest{\Stop} = \DeadEnd\;,
\\
\rest{\DeadEnd} = \DeadEnd\;,
\\
\rest{\pcc{t}{a}{t'}} = t\;,
\end{geqns}
\quad
\begin{geqns}
\resf{\Stop} = \DeadEnd\;,
\\
\resf{\DeadEnd} = \DeadEnd\;,
\\
\resf{\pcc{t}{a}{t'}} = t'\;.
\end{geqns}
\end{ldispl}

We write $\Bool^{\leq n}$, where $n \in \Nat$, for the set
$\set{u \in \seqof{\Bool} \where \len(u) \leq n}$.%
\footnote
{As usual, we write $\seqof{D}$ for the set of all finite sequences
 with elements from set $D$ and $\len(\sigma)$ for the length of finite
 sequence $\sigma$.
 Moreover, we write $\emptyseq$ for the empty sequence, $d$ for the
 sequence having $d$ as sole element, $\sigma \sigma'$ for the
 concatenation of finite sequences $\sigma$ and $\sigma'$, and
 $\tail(\sigma)$ for the tail of finite sequence $\sigma$.
}

It is assumed that a natural number $\maxlen$ has been given.
The number $\maxlen$ is taken for the maximal number of steps that the
instruction stream generator may run ahead of the instruction stream
execution unit.

The set $\Msg$ of \emph{instruction messages} is defined as follows:
\begin{ldispl}
\Msg = [0,\maxlen] \x \Bool^{\leq \maxlen} \x \BActi\;.
\end{ldispl}
In an instruction message $\tup{n,u,a} \in \Msg$:
\begin{itemize}
\item
$n$ is the number of replies that is acknowledged by the message;
\item
$u$ is the sequence of replies after which the instruction that is part
of the message must be executed;
\item
$a$ is the instruction that is part of the message.
\end{itemize}
The instruction stream generator sends instruction messages via an
instruction message transmission channel to the instruction stream
execution unit.
We refer to a succession of transmitted instruction messages as an
\emph{instruction stream}.
An instruction stream is dynamic by nature, in contradistinction with an
instruction sequence.

The set $\Stisg$ of \emph{instruction stream generator states} is
defined as follows:
\begin{ldispl}
\Stisg = [0,\maxlen] \x \setof{(\Bool^{\leq \maxlen+1} \x \Reg)}\;.
\end{ldispl}
In an instruction stream generator state $\tup{n,R} \in \Stisg$:
\begin{itemize}
\item
$n$ is the number of replies that has been received by the instruction
stream generator since the last acknowledgement of received replies;
\item
in each $\tup{u,p} \in R$, $u$ is the sequence of replies after which
the thread $p$ must be performed.
\end{itemize}
The functions $\updpm$ and $\updcr$ defined below are used to model the
updates of the instruction stream generator state on producing a message
and consuming a reply, respectively.
The function
$\funct{\updpm}{(\Bool^{\leq \maxlen} \x \Reg) \x \Stisg}{\Stisg}$
is defined as follows:
\begin{ldispl}
\updpm(\tup{u,p},\tup{n,R}) =
\\ \quad \left\{
\begin{array}{@{}l@{\quad}l@{}}
\tup{0,(R \diff \set{\tup{u,p}}) \union
       \set{\tup{u\True,\rest{p}},\tup{u\False,\resf{p}}}} &
\mif \act{p} \notin \set{\Stop, \DeadEnd}
\\
\tup{0,(R \diff \set{\tup{u,p}})} &
\mif \act{p} \in \set{\Stop, \DeadEnd}\;.
\end{array}
\right.
\end{ldispl}
The function $\funct{\updcr}{\Bool \x \Stisg}{\Stisg}$ is defined as
follows:
\begin{ldispl}
\updcr(e,\tup{n,R}) =
\tup{n + 1,\set{\tup{u,p} \where \tup{eu,p} \in R}}\;.
\end{ldispl}
The function $\select$ defined below is used to model the selection of
the sequence of replies and instruction that will be part of the next
message produced by the instruction stream generator.
The function
$\funct{\select}{\setof{(\Bool^{\leq \maxlen} \x \Reg)}}
                {\setof{(\Bool^{\leq \maxlen} \x \Reg)}}$
is defined as follows:
\begin{ldispl}
\select(R) =
\set{\tup{u,p} \in R \where
     \Forall{\tup{v,q} \in R}{\len(u) \leq \len(v)} \And
     \len(u) \leq \maxlen}\;.
\end{ldispl}
Notice that $\tup{u,p} \in \select(R)$ and $\tup{v,q} \in R$ only if
$u \leq v$.
By that depth-first run-ahead is excluded.
It happens that the performance of the protocol may change considerably
if the function $\select$ is  replaced by another function.

The set $\Stiseu$  of \emph{instruction stream execution unit states} is
defined as follows:
\begin{ldispl}
\Stiseu = [0,\maxlen] \x \setof{(\Bool^{\leq \maxlen} \x \BActi)}\;.
\end{ldispl}
In an instruction stream execution unit state $\tup{n,S} \in \Stiseu$:
\begin{itemize}
\item
$n$ is the number of replies for which the instruction stream
execution unit still has to receive an acknowledgement;
\item
in each $\tup{u,a} \in S$, $u$ is the sequence of replies after which
the action $a$ must be executed.
\end{itemize}
The functions $\updcm$ and $\updpr$ defined below are used to model the
updates of the instruction stream execution unit state on producing a
reply and consuming a message, respectively.
The function $\funct{\updcm}{\Msg \x \Stiseu}{\Stiseu}$ is defined as
follows:
\begin{ldispl}
\updcm(\tup{k,u,a},\tup{n,S}) =
\tup{n - k,S \union \set{\tup{\tail^{n - k}(u),a}}}\;.%
\footnotemark
\end{ldispl}
\footnotetext
{$\tail^n(u)$ is defined by induction on $n$ as usual:
 $\tail^0(u) = u$ and $\tail^{n+1}(u) = \tail(\tail^{n+1}(u))$.
}
The function $\funct{\updpr}{\Bool \x \Stiseu}{\Stiseu}$ is defined as
follows:
\begin{ldispl}
\updpr(e,\tup{n,S}) =
\tup{n + 1,\set{\tup{u,a} \where \tup{eu,a} \in S}}\;.
\end{ldispl}
The function $\enable$ defined below is used to distinguish between the
execution of a basic action $a \in \BAct$, which leads to a reply, and
the execution of $\Stop$ or $\DeadEnd$, which leads to termination or
inaction.
The function
$\funct{\enable}
  {\BActi \x \setof{(\Bool^{\leq \maxlen} \x \BActi)}}{\Bool}$
is defined as follows:
\begin{ldispl}
\enable(a,S) =
\left\{
\begin{array}{@{}l@{\quad}l@{}}
\True  & \mif \tup{\emptyseq,a} \in S
\\
\False & \mif \tup{\emptyseq,a} \notin S\;.
\end{array}
\right.
\end{ldispl}

Notice that the set $\Bool = \set{\True,\False}$ is the set of replies.
The instruction stream execution unit sends replies via a reply
transmission channel to the instruction stream generator.
We refer to a succession of replies as a \emph{reply stream}.

For the purpose of describing the transmission protocol in \ACPt, $\Act$
and $\commm$ are taken such that, in addition to the conditions
mentioned at the beginning of Section~\ref{sect-process-extr}, the
following conditions are satisfied:
\begin{ldispl}
\begin{aeqns}
\Act & \supseteq &
\set{\snd_i(d) \where i \in \set{1,2}, d \in \Msg} \union
\set{\rcv_i(d) \where i \in \set{1,2}, d \in \Msg}
\\ & {} \union {} &
\set{\snd_i(e) \where i \in \set{3,4}, e \in \Bool} \union
\set{\rcv_i(e) \where i \in \set{3,4}, e \in \Bool} \union
\set{\jact}
\end{aeqns}
\end{ldispl}%
and for all $i \in \set{1,2}$, $j \in \set{3,4}$, $d \in \Msg$,
$e \in \Bool$, and $e \in \Act$:
\begin{ldispl}
\begin{aeqns}
\snd_i(d) \commm \rcv_i(d) = \jact \;,
\\
\snd_i(d) \commm e = \dead & & \mif e \neq \rcv_i(d)\;,
\\
e \commm \rcv_i(d) = \dead & & \mif e \neq \snd_i(d)\;,
\eqnsep
\jact \commm e = \dead\;.
\end{aeqns}
\qquad\;
\begin{aeqns}
\snd_j(e) \commm \rcv_j(e) = \jact \;,
\\
\snd_j(e) \commm e = \dead & & \mif e \neq \rcv_j(e)\;,
\\
e \commm \rcv_j(e) = \dead & & \mif e \neq \snd_j(e)\;,
\end{aeqns}
\end{ldispl}

Let $p \in \Reg$.
Then the process representing the basic protocol for instruction stream
processing with regard to thread $p$ is described by
\begin{ldispl}
\encap{H}(\ISG{p} \parc \CHM \parc \CHR \parc \ISEU{})\;,
\end{ldispl}
where the process $\ISG{p}$ is recursively specified by the following
equations:
\begin{ldispl}
\renewcommand{\arraystretch}{1.5}
\begin{aeqns}
\ISG{p} & = & \ISGi{\tup{0,\set{\tup{\emptyseq,p}}}}\;,
\eqnsep
\ISGi{\tup{n,R}} & = &
\Altc{\tup{u,p} \in \select(R)}
 \snd_1(\tup{n,u,\act{p}}) \seqc \ISGi{\updpm(\tup{u,p},\tup{n,R})}
\\ & \altc &
\xAltc{e \in \Bool}{\tup{u,p} \in \select(R)}
 \rcv_4(e)  \seqc \ISGi{\updcr(e,\tup{n,R})}
\\
\multicolumn{3}{@{}l@{}}
{(\mathrm{for\; every\;} \tup{n,R} \in \Stisg
  \mathrm{\;with\;} R \neq \emptyset)\;,}
\eqnsep
\ISGi{\tup{n,\emptyset}} & = & \jact
\\
\multicolumn{3}{@{}l@{}}
{(\mathrm{for\; every\;} \tup{n,\emptyset} \in \Stisg)\;,}
\end{aeqns}
\end{ldispl}
the process $\CHM$ is recursively specified by the following equation:
\begin{ldispl}
\begin{aeqns}
\CHM & = &
\Altc{d \in \Msg}
 \rcv_1(d) \seqc \snd_2(d) \seqc \CHM\;,
\end{aeqns}
\end{ldispl}
the process $\CHR$ is recursively specified by the following equation:
\begin{ldispl}
\begin{aeqns}
\CHR & = &
\xAltc{e \in \Bool}{d \in \Msg}
 \rcv_3(e) \seqc \snd_4(e) \seqc \CHR\;,
\end{aeqns}
\end{ldispl}
the process $\ISEU{}$ is recursively specified by the following
equations:
\begin{ldispl}
\begin{aeqns}
\ISEU{} & = & \ISEUi{\tup{0,\emptyset}}\;,
\eqnsep
\ISEUi{\tup{n,S}} & = &
\xAltc{d \in \Msg}{f.m \in \BAct}
 \rcv_2(d) \seqc \ISEUi{\updcm(d,\tup{n,S})}
\\ & \altc &
\Altc{f.m \in \BAct}
 \enable(f.m,S) \gc \snd_f(m) \seqc \ISEUii{\tup{n,S}}
\\ & \altc &
\enable(\stopd,S) \gc \stp \altc
\enable(\deadd,S) \gc \iact \seqc \dead
\\
\multicolumn{3}{@{}l@{}}
{(\mathrm{for\; every\;} \tup{n,S} \in \Stiseu)\;,}
\eqnsep
\ISEUii{\tup{n,S}} & = &
\xAltc{e \in \Bool}{d \in \Msg}
 \rcv_f(e) \seqc \snd_3(e) \seqc \ISEUi{\updpr(e,\tup{n,S})}
\\ & \altc &
\Altc{d \in \Msg}
 \rcv_2(d) \seqc \ISEUii{\updcm(d,\tup{n,S})}
\\
\multicolumn{3}{@{}l@{}}
{(\mathrm{for\; every\;} \tup{n,S} \in \Stiseu)\;,}
\end{aeqns}
\end{ldispl}
and
\begin{ldispl}
\begin{aeqns}
H & = &
\set{\snd_i(d) \where i \in \set{1,2}, d \in \Msg} \union
\set{\rcv_i(d) \where i \in \set{1,2}, d \in \Msg}
\\ & {} \union {} &
\set{\snd_i(e) \where i \in \set{3,4}, e \in \Bool} \union
\set{\rcv_i(e) \where i \in \set{3,4}, e \in \Bool}\;.
\end{aeqns}
\end{ldispl}
$\ISG{p}$ is the instruction stream generator generating an instruction
sequence for thread $p$, $\CHM$ is the instruction message transmission
channel, $\CHR$ is the reply transmission channel, and $\ISEU{}$ is the
instruction stream execution unit.

The protocol described above has been designed so as to satisfy the
following equation:
\begin{ldispl}
\tau \seqc \pextr{p} =
\tau \seqc
\abstr{\set{\jact}}
 (\encap{H}(\ISG{p} \parc \CHM \parc \CHR \parc \ISEU{}))\;.
\end{ldispl}
We refrain from proving that the protocol satisfies this equation since
this paper is first and foremost a conceptual paper.

The transmission channels $\CHM$ and $\CHR$ can keep one instruction
message and one reply, respectively.
The protocol has been designed in such a way that the protocol will also
work properly if these channels are replaced by channels with larger
capacity and even by channels with unbounded capacity.

\section{Adaptations of the Protocol}
\label{sect-adaptations}

In this section, we discuss some conceivable adaptations of the protocol
described in Section~\ref{sect-protocol}.

Consider the case where, for each instruction, it is known what the
probability is with which its execution leads to the reply $\True$.
This might give reason to adapt the protocol described in
Section~\ref{sect-protocol}.
Suppose that the instruction stream generator states do not only keep
the sequences of replies after which threads must be performed, but also
the sequences of instructions involved in producing those sequences of
replies.
Then the probability with which the sequences of replies will happen can
be calculated and several conceivable adaptations of the protocol to
this probabilistic knowledge are possible by mere changes in the
selection of the sequence of replies and instruction that will be part
of the next instruction message produced by the instruction stream
generator.
Among those adaptations are:
\begin{itemize}
\item
restricting the instruction messages that are produced ahead to the ones
where the sequence of replies after which the instruction must be
executed will happen with a probability $\geq 0.50$, but sticking to
breadth-first run-ahead;
\item
restricting the instruction messages that are produced ahead to the ones
where the sequence of replies after which the instruction must be
executed will happen with a probability $\geq 0.95$, but not sticking to
breadth-first run-ahead.
\end{itemize}

Regular threads can be represented in such a way that it is effectively
decidable whether the two threads with which a thread may proceed after
performing its first action are identical.
Consider the case where threads are represented in the instruction
stream generator states in such a way.
Then the protocol can be adapted such that no duplication of instruction
messages takes place in the cases where the two threads with which a
thread possibly proceeds after performing its first action are
identical.
This can be accomplished by using sequences of elements from
$\Bool \union \set{*}$, instead of sequences of elements from $\Bool$,
in instruction messages, instruction stream generator states, and
instruction stream execution unit states.
The occurrence of $*$ at position $i$ in a sequence indicates that the
$i$th reply may be either $\True$ or $\False$.
The impact of this change on the updates of instruction stream generator
states and instruction stream execution unit states is minor.

\section{Conclusions}
\label{sect-concl}

We have described a basic protocol to deal with the phenomenon that, on
execution of an instruction sequence, a stream of instructions to be
processed arises at one place and the processing of that stream of
instructions is handled at another place.
By that we have brought this phenomenon better into the picture.
We have also discussed some conceivable adaptations of the basic
protocol.

The description of the protocol starts from the behaviours produced by
instruction sequences under execution.
By that we abstract from the instruction sequences which produce those
behaviours.
How instruction streams can be generated efficiently from instruction
sequences is a matter that obviously requires investigations at a less
abstract level.
The investigations in question are an option for future work.

We believe that the protocol described in this paper provides a setting
in which basic techniques aimed at increasing processor performance,
such as pre-fetching and branch-prediction, can be studied at a more
abstract level than usual (cf.~\cite{HP03a}).
In particular, we think that the protocol can serve as a starting-point
for the development of a model with which trade-offs encountered in the
design of processor architectures can be clarified.
We consider investigations into this matter an interesting option for
future work.

\bibliographystyle{spmpsci}
\bibliography{TA}


\end{document}